\documentstyle[psfig,gcdv]{article}
\def\reference{\parskip 0pt\par\noindent\hangindent 0.5 truecm}

\def\spose#1{\hbox to 0pt{#1\hss}}
\def\simlt{\mathrel{\spose{\lower 3pt\hbox{$\mathchar"218$}}
     \raise 2.0pt\hbox{$\mathchar"13C$}}}
\def\simgt{\mathrel{\spose{\lower 3pt\hbox{$\mathchar"218$}}
     \raise 2.0pt\hbox{$\mathchar"13E$}}}

\begin{document}

\small
\shorttitle{The Andromeda Stream}
\shortauthor{Lewis et al.}
%
%
\title{\large \bf
THE ANDROMEDA STREAM}

\author{\small
G. F. Lewis$^{1}$, R. A. Ibata$^{2}$, S. C. Chapman$^{3}$,
A. M. N. Ferguson$^{4}$, A. W. McConnachie$^{5}$,
M. J. Irwin$^{5}$ \& N. Tanvir$^{6}$
}

\date{}
\twocolumn[
\maketitle
\vspace{-20pt}
\small
{\center
$^1$ School of Physics, University of Sydney, Sydney, NSW, 2006\\
$^2$ Observatoire de Strasbourg, 11, rue de l'Universit\'e, F-67000, 
Strasbourg\\
$^{3}$
California Institute of Technology, Pasadena, CA 91125, U.S.A\\
$^4$ Max-Planck-Institut fur Astrophysik, Karl-Schwarzschild-Str. 1, 
Postfach 1317, D-85741, Garching, Germany\\
$^5$ Institute of Astronomy, Madingley Road, Cambridge, CB3 0HA, U.K.\\
$^6$ Physical Sciences, University of Hertfordshire, Hatfield, AL10 9AB,
U.K.\\ 
}

%
\begin{center}
{\bfseries Abstract}
\end{center}
\begin{quotation}
\begin{small}
\vspace{-5pt}
The  existence  of  a  stream  of  tidally  stripped  stars  from  the
Sagittarius Dwarf galaxy  demonstrates that the Milky Way  is still in
the process of accreting mass.   More recently, an extensive stream of
stars has  been uncovered in the  halo of the  Andromeda galaxy (M31),
revealing that it too is  cannibalizing a small companion.  This paper
reports the recent observations of this stream, determining it spatial
and kinematic properties, and tracing its three-dimensional structure,
as well as describing future  observations and what we may learn about
the Andromeda galaxy from this giant tidal stream.
\\
{\bf Keywords: 
Galaxy: kinematics and dynamics ---
methods: N-body simulations
}
\end{small}
\end{quotation}
]


\bigskip

\section{Introduction}
${\rm  \Lambda}$CDM has  become  the preferred  model of  cosmological
structure formation,  providing a convincing description  of the large
scale  distribution of matter  in the  Universe.  On  galactic scales,
however, this paradigm  has proved somewhat unsatisfactory, predicting
a myriad of satellite systems  accompanying the Milky Way which do not
appear to be there (Klypin et  al. 1999). One clear prediction of this
model, however,  is that large galaxies  like the Milky  Way grew over
time via the accretion of smaller systems.

Numerical simulations of such  accretion events reveal that they leave
long   lived  {\it  fossil}   signatures  within   the  halo   of  the
cannibalizing galaxy in  the form of extensive tidal  streams that can
completely  wrap the host  (e.g.  Johnston  1998).  The  detection and
characterisation  of  this  fossil  record  will  unravel  the  recent
accretion history of a galaxy.   While there is some evidence that the
Magellanic  Clouds   may  represent  a  case   of  ongoing  accretion,
displaying  an extensive gaseous  tail, the  first clear  detection of
current cannibalization  in the  halo of the  Milky Way came  with the
discovery of the  Sagittarius Dwarf galaxy in 1994  (Ibata, Gilmore \&
Irwin 1994).  Since its  initial detection, studies have revealed more
and more stellar debris located farther and farther from the main body
of the dwarf (e.g. Majewski et  al. 1999) .  Ibata et al. (2001) found
a stream  of carbon stars lying over  a great circle on  the sky which
intersects  with  the  current  location  of  the  Sagittarius  Dwarf.
Furthermore,  this stream  is also  aligned with  Sagittarius's proper
motion, clearly  demonstrating it represents  material associated with
the dwarf  galaxy.  Intriguingly, the collimated nature  of the stream
strongly suggests that the dark matter halo is spherical, at odds with
theoretical   expectations   of   a   strongly   flattened,   triaxial
distribution.  These  results were  recently confirmed by  Majewski et
al.  (2003) using a larger sample of stars drawn from the 2-Micron All
Sky Survey (2MASS).

While  the  study of  Sagittarius  has  proved  quite successful,  our
location within the Milky Way limits our observational prospects, with
any tidal  debris presenting an  extremely low stellar density  on the
sky. Furthermore, it is very important to know if the situation of the
Milky Way  is unique or if  its current appetite  is representative of
galaxies in general. We need,  therefore, to turn our attention to the
search for tidal features  in external galaxies.  While this increases
the apparent  stellar density of tidal debris,  distance rapidly blurs
individual  stars  into   uniform  surface  brightness  and  dynamical
measures become extremely time  consuming. Hence, the search for tidal
debris should  be aimed  at our  nearest companions if  we are  to use
their structural and kinematic properties in a fashion similar to that
of the Sagittarius stream.

\section{The Andromeda Stream}\label{stream}

\subsection{Wide Field Photometry}
M31,  being  our  extragalactic  neighbour,  has been  the  target  of
numerous observational programs.  With the advent of CCDs earlier work
focused  on  deep  pencil-beam   studies  which  indicated  a  complex
metallicity  mix  in M31's  halo.   In  contrast,  this present  study
employed the Wide Field Camera on the Isaac Newton Telescope to obtain
a deep but panoramic survey of  the stellar populations in the halo of
M31.  Covering  0.3 deg$^2$  per pointing, the  initial survey  in the
year 2000 tiled a region out  to 4$^o$ (55kpc) of the Southern portion
of the  halo of M31, covering  an area of  $\sim$10deg$^2$ to $i=23.5$
and $V=24.5$ (Ibata et al. 2001).

An examination of the halo stellar density by eye clearly reveals that
it is  not smooth,  showing significant substructure,  particularly an
apparent stream of  stars stretching to the south of  the main body of
Andromeda.  This  feature is significantly enhanced if  a selection is
made of metal  rich red giant branch (RGB) stars  (see Figure 1; note,
this figure  contains further observations  with the INT  WFC, mapping
out the  northern sector of the  halo also). As well  as the prominent
tidal stream, these data also  reveal complex structure in the halo of
M31, including  the northern  spur and a  significant over  density of
stars in  the vicinity of the  giant globular cluster  G1 (Ferguson et
al.  2002).  The  extensive substructure  suggests that  M31  may have
undergone  more  recent  accretion  events, resulting  in  the  rather
complex metallicity distribution within its halo.

\begin{figure}
\begin{center}
\begin{tabular}{c}
\psfig{file=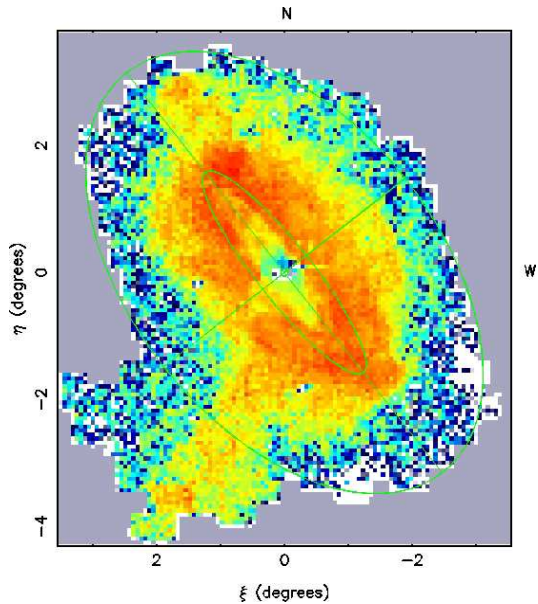,width=7cm}
\end{tabular}
\caption{RGB stars as  derived from the INT WFC  survey. The Andromeda
Stream of stars is clearly visible extending to the south.  }
\label{figure1}
\end{center}
\end{figure}

\subsection{The distance to the Stream}
In unraveling the history of  the stream, it is important to determine
its three-dimensional structure. To  this end, the stream was targeted
with   the  12K  camera   on  the   Canada  France   Hawaii  Telescope
(CFH12K). Covering a total of $\sim$3deg$^2$, 14 fields were obtained,
starting $\sim$5 degrees below the  plane of M31, and ending $\sim$2.5
degrees   above   the   plane   (Figure   2).    An   examination   of
colour-magnitude  diagrams of  the  CFH12K fields  clearly reveal  the
presence of  the tidal stream  of stars as  a distinct feature  to the
very  southern extremity of  the survey.   The situation  is, however,
different  in the  north where  the  signature of  the stream  quickly
peters out, disappearing in  the northernmost field; this reflects the
structure seen in the INT imaging survey described previously.

\begin{figure}
\begin{center}
\begin{tabular}{c}
\psfig{file=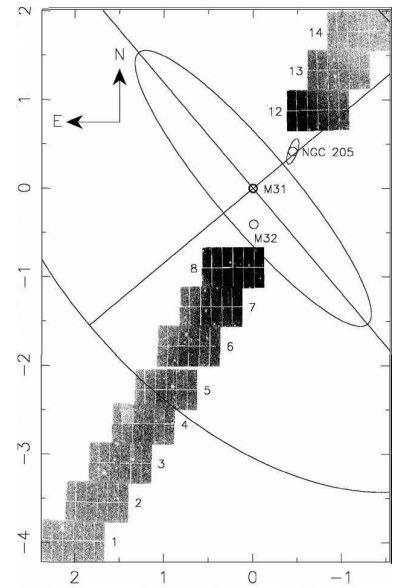,width=5cm}
\end{tabular}
\caption{The location of  the CFHT 12K fields relative  to the central
regions  of M31.  Three  fields crossing  the plane  of M31  have been
removed as  they were not employed  in the analysis  of McConnachie et
al. (2003) due to severe crowding issues.}
\label{figure2}
\end{center}
\end{figure}

The  superb   quality  of  the  CFH12K  data,   however,  provided  an
opportunity to measure the distance to the stream at various locations
along its length.  This was achieved by first determining the location
of the tip of the red giant branch (TRGB) in the main body of M31 at a
range of metallicities.  By comparing  the location of the TRGB in the
fields along  the stream with  the main body  of M31, the  position of
each field relative  to M31 could be measured  via a cross-correlation
of the stellar luminosity functions, as any offsets would primarily be
due  to a  difference  in  distance (McConnachie  et  al. 2003).   The
results of  this study are presented  in Figure 3,  revealing that the
Andromeda stream curves away from us  below the disk of M31.  The most
southern extremity  of the survey  lies at a distance  of $\sim$900kpc
from us, 120kpc farther than M31 itself.  In the North, the fields are
actually in front of M31, but are curving away from us northwards.

\subsection{Stellar Velocities}
While the spatial structure  of the Andromeda Stream reveals important
clues  to its  origin and  evolution, fully  unraveling  its dynamical
history requires a measurement of  the stellar kinematics. As the wide
field  imaging  has resolved  the  stream  into  individual stars  the
determination  of  stellar  velocities  is  possible  with  8-m  class
telescopes.   To this end,  a series  of observations  were undertaken
using the  DEIMOS spectrograph on the 10-m  Keck2 telescope. Employing
multislit masks, resulting with $\sim100$ spectra over a $16'\times5'$
region, four  fields along the  stream were targeted. The  RGB sources
possessed $20.5<i<22.0$, resulting in a velocity accuracy of 5-10km/s.

These  observations  revealed a  strong  velocity  gradient along  the
stream, with the southernmost region of the stream traveling toward us
at the  systemic velocity of M31, while  in the vicinity of  M31 it is
approaching  at  300km/s with  respect  to  M31  (Ibata et  al.,  {\it
submitted}).  Coupled with the distance determinations outlined above,
these kinematics provide strong  constraints on the orbital properties
of  the stream;  Figure~\ref{figure4} presents  two  preliminary orbit
fits to the extant kinematic and spatial data of stars in the stream.

\begin{figure}
\begin{center}
\begin{tabular}{c}
\psfig{file=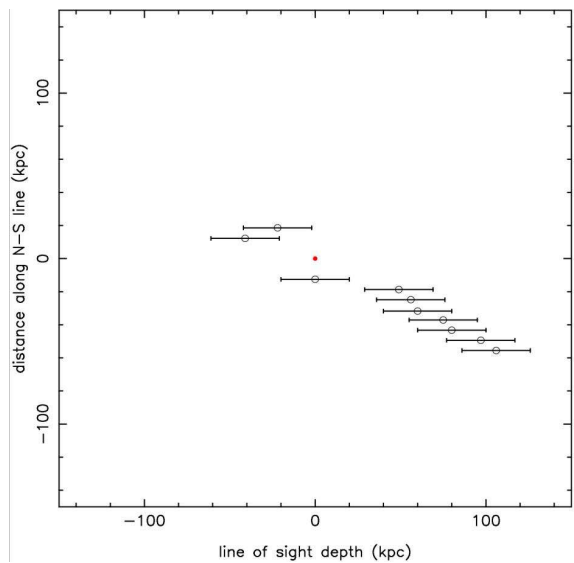,width=7.5cm}
\end{tabular}
\caption{The  distance  to the  Andromeda  Stream  as determined  from
calibrating the tip of the  red giant branch and cross-correlating the
luminosity function of stars along the Andromeda Stream.}
\label{figure3}
\end{center}
\end{figure}

\subsection{Interpretation}\label{Interpretation}
The stream orbits allow us  constrain the mass distribution in the M31
halo. A  full analysis will  require N-body simulations to  be carried
out in  different halo potentials,  with the resulting  stellar stream
compared  to  the  observed  position,  distance  and  velocity  data.
However, a good approximation can  be obtained by simply comparing the
locus and  velocity profile of test  particle orbits, a  task which is
computationally  much cheaper.   For this  approximation to  work, the
dwarf galaxy progenitor  of the stream must be  of relatively low mass
$<  10^9  M_\odot$,   for  the  self-gravity  of  the   stream  to  be
unimportant. By comparing the best-fit orbit to the data as a function
of halo potential,  the most likely value of the total  mass of M31 is
$M_{tot} = 6.5  \pm 1.6 \times 10^{11} M_\odot$  out to $145kpc$. With
this,  Andromeda  has a  very  similar halo  mass  to  the Milky  Way.
However, this  result is somewhat  uncertain; using only  the southern
data to contstrain the form of  the potential reduces the halo mass by
a  factor of  $\sim4$ [see  Ibata et  al. ({\it  Submitted})  for more
details].

\section{Future Observations}\label{Future}
In  September  2003, further  DEIMOS  spectroscopy  was undertaken  at
Keck2.   To increase  the observational  multiplexing to  roughly 1000
stars per mask, a narrow-band  filter was employed to capture only the
region of the CaII NIR triplet.  In this manner we targeted a total of
$\sim2\times10^4$  stars.   This  data  will be  supplemented  with  a
CFHT/MEGACAM survey of the southern  quadrant of M31, covering a total
of  almost 80  square degrees,  calibrating the  entire extent  of the
tidal stream and allowing us to probe the halo between M31 and M33.

\begin{figure}
\begin{center}
\begin{tabular}{c}
\psfig{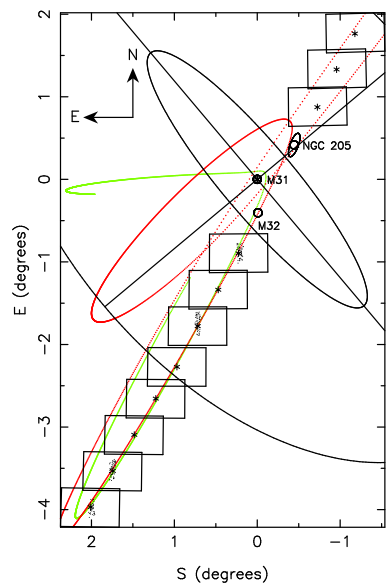}
\end{tabular}
\caption{Two preliminary  fits to the  orbit of the  Andromeda stream.
The green  model utilises only the  southern data; the  overall fit to
the  kinematic and spatial  data is  good but  the resulting  orbit is
strongly radial.   The red model, incorportating the  entire data set,
is a  poor representation of the data, but its
orbit is  similar to the rosette  orbit of the  Sagittarius dwarf.}
\label{figure4}
\end{center}
\end{figure}

This extensive dataset will allow a more complete determination of the
orbital  properties  of  the  Andromeda stream  and  characterise  its
dynamical evolution and  ultimate demise.  As with the  studies of the
Sagittarius tidal stream (e.g. Ibata  et al. 2001), this approach will
reveal the form of the dark matter halo of M31.

\section*{Acknowledgments}
GFL thanks  Swinburne University's Astronomy  Group for hosting  a fun
and informative meeting, although Sydney would have been warmer.

\section*{References}

\reference Ferguson, A.~M.~N., 
Irwin, M.~J., Ibata, R.~A., Lewis, G.~F., \& Tanvir, N.~R.\ 2002, AJ, 124, 
1452 

\reference Ibata, R.~A., 
Gilmore, G., \& Irwin, M.~J.\ 1994, Nature, 370, 194 

\reference Ibata, R., Lewis, G.~F., 
Irwin, M., Totten, E., \& Quinn, T.\ 2001, ApJ, 551, 294 

\reference Ibata, R., Irwin, M., Lewis, G., Ferguson, A.~M.~N., \& 
Tanvir, N.\ 2001, Nature, 412, 49 

\reference Johnston, K.~V.\ 1998, ApJ, 495, 297 

\reference Klypin A., Kravtsov A.~V., Valenzuela O., Prada F.\ 1999, 
ApJ, 522, 82 

\reference Majewski, S.~R., 
Siegel, M.~H., Kunkel, W.~E., Reid, I.~N., Johnston, K.~V., Thompson, 
I.~B., Landolt, A.~U., \& Palma, C.\ 1999, AJ, 118, 1709 

\reference Majewski, S.~R., Skrutskie, M.~F., Weinberg, M.~D. \& 
Ostheimer, J.~C., 2003, {\it astro-ph/0304198}

\reference McConnachie, A.~W., 
Irwin, M.~J., Ibata, R.~A., Ferguson, A.~M.~N., Lewis, G.~F., \& Tanvir, 
N.\ 2003, MNRAS, 343, 1335 

\end{document}